\documentstyle{elsart}
\begin{document}

Quantum Field Theory by Lewis H. Ryder, Cambridge University Press, 487 pages,
Second Edition (1996).

D. V. Ahluwalia (Los Alamos National Laboratory).

The first edition of this book appeared in 1985. To my knowledge this
was the the first book that presented a completely modern and logical
development of quantum theory of matter and gauge fields.
Since this book had very significant effect on my research,
the review that follows carries the flavor of a personal note of
thanks.

So in late eighties, when the first edition of this book reached my desk,
I was a graduate student
who had just 
finished taking some elementary courses in quantum mechanics
and quantum field theory. I had not yet chosen a research topic.
Professor Lewis H. Ryder's relatively slim book arrived my desk,
and I, who was suffering from intense frustration created by the lack
of a modern classic in quantum field theory, was immediately enamored
by this work of scholarship.

The initial chapters of
Professor Ryder's slim monograph guided me to the original works of
Eugene Wigner, Steven Weinberg, C. N. Yang and R. L. Mills. 
Within a matter of a year or so I had expanded some of Professor
Ryder's work and finished a over--two--hundred 
page Ph.D dissertation under my friend
and mentor Professor Dave Ernst, 
a nuclear physicist who was at that time interested 
in high--spin hadronic resonances.

In the first chapter the reader is be introduced to the
a ``synopsis of particle physics,''
with the request, ``I ask readers' indulgence to make the best they can
of these sections until they meet explanations later in the book.''

Chapter 2 is where the real fun begins. It is devoted to
``single--particle relativistic wave equation.'' Here one need not depend
on Dirac's genius to obtain the celebrated spin-1/2 wave equation for
charged particles. Elegant and simple arguments based on classic works of
Wigner, Weinberg, and his own teacher Feza G\"ursey, are exploited
to {\em derive} the  celebrated Dirac equation. In the second edition the 
author had the opportunity to {\em correct} an important element that enters the 
derivation of Dirac equation. On. p. 41 (p.44 of the first edition) 
one is taught, ``Now, when a particle is at rest, one cannot define its 
spin as either left--or right--handed, so $\phi_R(0)\,=\,\phi_L(0)$.''
This important observations should have been corrected, as I learned with
C. Burgard,  to read:
\begin{quote}
Now, when a particle is at rest, one cannot define its 
spin as either left--or right--handed, so $\phi_R(0)\,=\pm \,\phi_L(0)$.
\end{quote}
The $\pm$ sign is important for a consistent and correct derivation of the
Dirac equation and is responsible for  the {\em opposite}
 relative intrinsic parities of
spin--$1/2$ fermions. Its deeper significance arises when one considers spin one.
For spin one, the  same arguments (when corrected, as above) 
as that of Professor  Ryder lead to a fundamentally new construct.
Similarly, I wonder why after having so beautifully derived
the Dirac equation, Professor Ryder chose not to present the derivation
of the Maxwell equations as well. 
Within the framework of chapter 2, one could have easily obtained
the Dirac spinors {\em without} any reference to Dirac equation.
Again, this is not done and an opportunity is lost in showing the 
strength and beauty of Ryder's approach.

But these criticisms would not
have even arisen had Professor Ryder not shown me the path on which these 
questions and my subsequent researches lay!

The free wave equations have no fire. The fire is breathed by 
demanding their covariance under appropriate gauge group.
This is elegantly and logically done in Chapter 3.

The Chapter 2 and Chapter 3 provide the basic building blocks
for a quantum theory of interacting matter and gauge fields.
Chapter 4, as the logic demands, presents ``canonical quantisation and
particle interpretation.'' Chapter 5 is devoted to Richard Feynman's 
path integral approach. The Chapters 6 to 9 deal with the
standard perturbative approach of quantum field theory, including an
introduction to the standard model of electroweak interactions and
renormalisation. 

Chapter 10 and 11 are essentially self--contained essays on 
non-perturbative topological aspects and supersymmetry.
On a recent weekend, looking at the fallen foot and half of snow
in these mountains through a large glass window
framed by pine trees
in Los Alamos, I entertained myself with
parts of these essays.

A rare combination of a thorough understanding and appreciation 
of the essential logical structure of quantum field theory and deep
pedagogic skills have intermingled to create a masterpiece on the 
elementary introduction to quantum field theory in less than five 
hundred pages. My deepest thanks to Professor Ryder on behalf 
of all his readers.
Without reservations, I give my strongest recommendation to
every beginning student of physics to acquire and read 
{\em Quantum Field Theory} by L. H. Ryder.

\end{document}